\documentclass[aps,prl,twocolumn,groupedaddress]{revtex4}
\usepackage{graphicx}
\begin{document}

\title{Pairing symmetry and properties of iron-based high temperature superconductors}

\author{Yuan Wan}
\author{Qiang-Hua Wang}

\email[E-mail address:]{qhwang@nju.edu.cn}
\address{National Laboratory of Solid State Microstructures \& Department of Physics, Nanjing University,
Nanjing 210093, China}

\date{\today}

\begin{abstract}
Pairing symmetry is important to indentify the pairing mechanism.
The analysis becomes particularly timely and important for the
newly discovered iron-based multi-orbital superconductors. From
group theory point of view we classified all pairing matrices (in
the orbital space) that carry irreducible representations of the
system. The quasiparticle gap falls into three categories: full,
nodal and gapless. The nodal-gap states show conventional Volovik
effect even for on-site pairing. The gapless states are odd in
orbital space, have a negative superfluid density and are
therefore unstable. In connection to experiments we proposed
possible pairing states and implications for the pairing
mechanism.
\end{abstract}

\maketitle

\textit{Introduction} The newly discovered family of iron-based
ReOFeAs(Re = La, Ce, Pr, etc.) high temperature superconductors
are raising great interests in the community.\cite{material} The
superconductor consists of layers of FeAs which is believed to be
the conducting planes. The ReO layers in between the FeAs layers
stabilize the structure and donate carriers to the FeAs layers.
For example, substitution of O by F or introducing O
vacancies\cite{vacancy} dopes electrons into the system, and
contrarily substitution of Re by alkaline-earth elements can
realize hole doping.\cite{holedope} Some preliminary experimental
results on the superconducting (SC) state have been obtained. The
specific capacity measurement\cite{specific heat} and nuclear
magnetic resonance (NMR) \cite{NMR} indicate a nodal gap, and the
existence of Andreev bound state implies a non-trivial phase
structure of SC gap\cite{Andreev}, while both s-wave and dirty
d-wave behaviors are suggested by penetrate length
measurement\cite{superfluid}. On the theoretical side,
local-density-approximation (LDA) calculations show that the
states near the Fermi level are largely contributed by the five
d-orbitals. \cite{dorbital} Moreover, some authors demonstrated
that it is sufficient to keep only a few of them, for example the
$d_{xz}$ and $d_{yz}$ orbitals, to reproduce qualitatively the LDA
Fermi surface topology.\cite{twoorbital} As for the driving force
for superconducting pairing, both electronic\cite{electron} and
phonon-mediated mechanisms\cite{phonon} are proposed, and various
pairing symmetries are predicted.

Since the pairing symmetry is related to the pairing mechanism, a
classification of all possible pairing symmetries \cite{symmetry}
is important. This is more so given the fact that the pairing
function becomes an orbital-wise matrix function in the
multi-orbital case, which we elaborate in this paper. The main
results are as follows. 1) From a two-orbital ($d_{xz}+d_{yz}$)
model we classify all possible on-site and bond-wise pairing basis
matrices. In addition to the momentum dependence, the matrices
themselves carry nontrivial symmetries, so that even on-site
pairing may lead to a nodal or gapless pairing. 2) In the gapless
case the density of states (DOS) at the fermi level is enhanced by
the SC order. 3) Only the nodal-gap cases show Volovik effect in
an applied magnetic field. 4) Most surprisingly, the gapless SC
state may have a negative superfluid density, and is therefore
unstable against phase twisting. 5) In connection to available
experiments we propose possible pairing bases and further
experiments to reduce the candidate list.

\textit{Symmetry Analysis} In the FeAs layer there are two Fe ions
per unit cell because of the As ions. In order to simplify the
analysis, we adopt a two-band model, i.e. we keep the two
atomically degenerate $d_{xz}$ and $d_{yz}$ orbitals of Fe ion,
which are important for superconductivity, and neglect $d_{xy}$,
$d_{x^2-y^2}$ and $d_{3z^2-r^2}$ orbitals of Fe ions and all
orbitals of As ions for a moment. Thus we will focus on the
Fe-lattice, for which there is only one Fe ion in each unit cell.
The effect of As ions can be partially included in the effective
hopping integrals for the d-orbitals. We define $x$- and
$y$-direction unitary vectors as connecting the nearest neighbor
Fe-atoms. The space group of our model is $P4/mmm$, which has
higher symmetry than the space group of whole system $P4/nmm$. The
normal state of the model can be described by, in the momentum
space, $H_0=\sum_{k\sigma} \phi_{k\sigma}^\dagger
\xi_k\phi_{k\sigma}$ where
$\phi_k^\dagger=(d_{xz}^\dagger,d_{yz}^\dagger)_{k\sigma}$, and
$\xi_k=\epsilon_k\tau_0+\delta_k\tau_3+\gamma_k\tau_1$ in which
$\tau_{0,1,3}$ are unit and Pauli matrices defined in the orbital
space. For point group operations under concern, the $d_{xz}$ and
$d_{yz}$ orbital wave functions transform as $x$ and $y$,
respectively. This dictates that $\phi_k^\dagger\tau_0\phi_k$
transforms as $A_{1g}$, $\phi_k^\dagger\tau_1\phi_k$ as $B_{2g}$,
$\phi_k^\dagger\tau_2\phi_k$ (which is actually absent in $H_0$
but is included here for later use) as $A_{2g}$, and
$\phi_k^\dagger\tau_3\phi_k$ as $B_{1g}$. Equivalently we claim
that the $\tau$-matrices carries the above-mentioned irreducible
representations, without referring to the orbital wave functions
further. Finally to leave $H_0$ invariant, $\epsilon_k$,
$\delta_k$ and $\gamma_k$ must transform as $A_{1g}$, $B_{1g}$ and
$B_{2g}$, respectively. The concrete form of the dispersions in
$H_0$ (see below) is irrelevant at this stage.

We now discuss the pairing symmetry. The system is invariant under
spin-SU(2) transformation and one can classify the pairing states
into spin-singlet and spin-triplet cases, which we assign in the
last step according to the global antisymmetry of the pairing
function with respect to the combined exchange of spin, orbital
and spatial position. We therefore concentrate first on the
symmetry of pairing as a function of momentum and a matrix in the
orbital space. Since the pairing operator
$\phi_{\alpha\sigma}(-k)\tau_i^{\alpha\beta}\phi_{\beta\sigma'}(k)$
(where $\phi_{\alpha,\beta}=d_{xz,yz}$ and $i=0,1,2,3$) transforms
identically as
$\phi^\dagger_{\alpha\sigma}(k)\tau_i^{\alpha\beta}\phi_{\beta\sigma'}(k)$,
we immediately see that the $\tau$-matrices in the pairing matrix
transform exactly as they do in $H_0$. For on-site pairing,
$\tau_{0,1,2,3}$ carries the irreducible representations. Since
$\tau_{0,1,3}$ are even in orbital space and transform as
$A_{1g}$, $B_{2g}$ and $B_{1g}$ under the point group, the spin
channel must be a singlet. On the other hand, $i\tau_2$ is odd in
orbital space and transform as $A_{2g}$, the spin channel must be
in a triplet.

The extension to pairing on bonds is almost straightforward. We
may multiply the above mentioned $\tau$-matrices by trigonometric
basis functions to form the pairing matrices corresponding to
pairing on bonds in real space. For the system under concern, the
basis functions for nearest-neighbor bonds are $\cos k_x+\cos k_y$
($A_{1g}$), $\cos k_x-\cos k_y$ ($B_{1g}$), $(\sin k_x,\sin k_y)$
(carrying the two-dimensional $E_g$ representation). The symmetry
of the product is easily seen from that fact that the
$\tau$-matrices only carry one-dimensional representations. For
example, $B_{1g}\cdot \tau_1\sim B_{1g}\cdot B_{2g}\sim A_{2g}$.
Since the orbital parity is even, the spin channel must be a
singlet. In Tab.I we list all possible pairing basis matrices for
on-site pairing and nearest-neighbor-bond pairing (extension to
longer-range pairing is trivial), together with the irreducible
representations they carry, the spin symmetry, orbital parity, and
the behavior of the corresponding quasiparticle excitation gap in
the momentum space (which will be discussed later). We notice that
even on-site pairing can carry a non-trivial representation,
indicating the unique role of the two (atomically) degenerate
d-wave orbitals.

\begin{table}
\caption{\label{symmetry} Pairing basis matrices carrying
irreducible representations of the model. The first column is the
index number, the second and the third columns list the
representations and the basis matrix functions. The spin and
orbital parities are shown in the forth and the fifth columns. The
last column describe the behavior of quasiparticle excitation gap
in the momentum space. Notice that $\tau_{0,1,2,3}$ transform as
$A_{1g}, B_{2g}, A_{2g},B_{1g}$ respectively due to two $xz$ and
$yz$ orbitals. These rules are important to identify the global
symmetry and the gap behavior of the basis matrix function.}
\begin{ruledtabular}
\begin{tabular}{ccccccc}
No. & IR & Basis & Spin & Orbital Parity & Gap\\
\hline
1 & $A_{1g}$ & $\tau_{0}$ & S & E &  Full \\
2 & $A_{1g}$ & $(\cos k_x+\cos k_y)\tau_{0}$ & S & E & Nodal \\
3 & $A_{1g}$ & $(\cos k_x-\cos k_y)\tau_{3}$ & S & E &  Nodal \\
4 & $A_{2g}$ & $(\cos k_x-\cos k_y)\tau_{1}$ & S & E &  Nodal \\
5 & $B_{1g}$ & $\tau_{3}$ & S & E &  Nodal \\
6 & $B_{1g}$ & $(\cos k_x-\cos k_y)\tau_{0}$ & S & E &  Nodal \\
7 & $B_{1g}$ & $(\cos k_x+\cos k_y)\tau_{3}$ & S & E &  Nodal \\
8 & $B_{2g}$ & $\tau_{1}$ & S & E &  Nodal \\
9 & $B_{2g}$ & $(\cos k_x+\cos k_y)\tau_{1}$ & S & E &  Nodal \\
10 & $E_{g}$ & $\left\{\begin{array}{c}\sin k_x i\tau_{2}\\{}\sin k_y i\tau_{2}\end{array}\right.$ & S & O  & Gapless \\
10' & $E_{g}$ & $\left\{\begin{array}{c}(\sin k_x+i\sin k_y)i\tau_{2}\\(\sin k_x-i\sin k_y)i\tau_{2}\end{array}\right.$ & S & O & Gapless \\
11 & $A_{2g}$ & $i\tau_{2}$ & T & O & Gapless \\
12 & $A_{2g}$ & $(\cos k_x+\cos k_y)i\tau_{2}$ & T & O &  Gapless \\
13 & $B_{2g}$ & $(\cos k_x-\cos k_y)i\tau_{2}$ & T & O &  Gapless \\
14 & $E_{g}$ & $\left\{\begin{array}{c}\sin k_x \tau_{0}\\{}\sin k_y\tau_{0}\end{array}\right.$ & T & E &  Nodal \\
14' & $E_{g}$ & $\left\{\begin{array}{c}(\sin k_x+i\sin k_y)\tau_{0}\\(\sin k_x-i\sin k_y)\tau_{0}\end{array}\right.$ & T & E  & Full \\
15 & $E_{g}$ & $\left\{\begin{array}{c}\sin k_x \tau_{3}\\{}\sin k_y\tau_{3}\end{array}\right.$ & T & E &  Nodal \\
15' & $E_{g}$ & $\left\{\begin{array}{c}(\sin k_x+i\sin k_y)\tau_{3}\\(\sin k_x-i\sin k_y)\tau_{3}\end{array}\right.$ & T & E  & Nodal \\
16 & $E_{g}$ & $\left\{\begin{array}{c}\sin k_x\tau_{1}\\{}\sin k_y\tau_{1}\end{array}\right.$ & T & E &  Nodal \\
16' & $E_{g}$ & $\left\{\begin{array}{c}(\sin k_x+i\sin k_y)\tau_{1}\\(\sin k_x-i\sin k_y)\tau_{1}\end{array}\right.$ & T & E  & Nodal \\
\end{tabular}
\end{ruledtabular}
\end{table}

\begin{figure}
\includegraphics[width=0.5\textwidth]{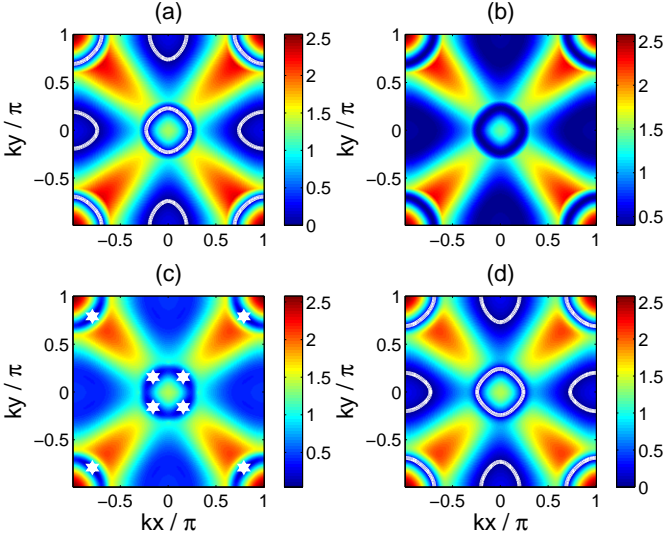}
\caption{\label{gap}(Color) Minimal quasiparticle excitation
energy as a function of momentm in the normal state (a), a
full-gap SC state (b), a nodal-gap SC state (c) and a gapless SC
state (d). Thick white lines (symbols) indicate the Fermi surface
(nodal points).}
\end{figure}
\textit{Quasiparticle excitation gap} To illustrate the concrete
gap structure in the various SC states, we need the band structure
of iron-based superconductors. Here we adopt the tight-binding
model introduced by Ref.\cite{band}. The BdG Hamiltonian is given
by $H=\sum_k \psi^\dagger_k H_k\psi_k$, where we recall that the
momentum is defined in the "large" Brillouin (BZ) corresponding to
one Fe per unit cell, $\psi_k=(d_{xz,k\uparrow}, d_{yz,k\uparrow},
d^{\dagger}_{xz, -k\downarrow}, d^{\dagger}_{yz,
-k\downarrow})^{T}$ is the four-component spinor in the orbital
and Nambu space, and $H_k$ is, in a form of block-matrix,
\begin{eqnarray*}
H_k=\left(\begin{array}{cc} {\xi_k} & {V\Delta_k}\\
{V\Delta^\dagger_k} & {-\xi_k}
\end{array}\right).
\end{eqnarray*}
Here $\xi_k=\epsilon_k\tau_0+\delta_k\tau_3+\gamma_k\tau_1$ is a
tight-binding dispersion defined in $H_0$, with $ \epsilon_k=
-(t_1+t_2)(\cos k_x+\cos k_y)-4t_3\cos k_x \cos k_y-\mu$,
$\delta_k  = -(t_1-t_2)(\cos k_x-\cos k_y)$, and $\gamma_k =
-4t_4\sin k_x \sin k_y$. Here $t_1=-1$, $t_2=-1.3$,
$t_3=t_4=-0.85$, and $\mu=1.45$, in unit of $|t_1|$. The gap
amplitude $V=0.4$ is chosen for illustration, and $\Delta_k$ is
selected from the basis matrix functions listed in Tab.I. The
hamiltonian can be exactly diagonalized to obtain the
quasiparticle excitations and other SC properties. Zero energy
excitations are determined by $\det(H_k) =0$. In the case of SC
state No.1 in Tab.I,
$\det(H_k)=(\epsilon_k^2-\delta_k^2-\gamma_k^2-V^2)^2+(2V\epsilon_k)^2=0$
has no solution. This corresponds to the full-gapped case. For SC
state No.5,
$\det(H_k)=(\epsilon_k^2-\delta_k^2-\gamma_k^2+V^2)^2+(2V\delta_k)^2=0$
is satisfied at four sets of nodal points in BZ. Another
interesting example is SC state No.11, for which
$\det(H_k)=(\epsilon_k^2-\delta_k^2-\gamma_k^2+V^2)^2=0$ holds
along lines in BZ. Such zero-lines form the fermi surface (FS)
(slightly different from the normal state FS) for the BdG
quasiparticles. This is the gapless case.

Due to the two-orbital character of the SC state, the nodal points
or the BdG FS may be located away from the normal state FS. We
therefore need the quasiparticle excitation energy in the whole BZ
to characterize the gap structure. Several typical results are
shown in Fig.\ref{gap} where the minimal quasiparticle excitation
energy as a function of momentum in (a) the normal state, (b) a
full-gap SC state No.1, (c) a nodal-gap SC state No.5 and (d) a
gapless SC state No.11. Here thick white lines (symbols) highlight
the zero excitation energy contour (nodal points). By checking the
excitation spectra of all possible cases in Tab.I, we summarize
that: (i) All $\tau_{1,3}$-bases are nodal, consistent with the
fact that these $\tau$ matrices carry $B_{2g,1g}$ representations.
(ii) All $\tau_2$-bases (which carries odd orbital-parity) are
gapless (unless the pairing energy scale is of the order of the
band width). (iii) The $\tau_0$-bases leads to full or nodal gaps,
depending on the momentum basis function.

\begin{figure}
\includegraphics[width=0.5\textwidth]{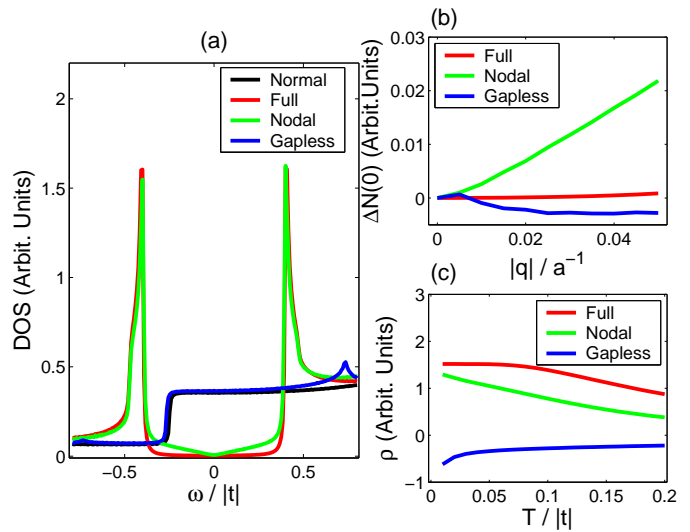}
\caption{\label{dos}(Color) Physical properties of various SC
states: (a)Density of states; (b) Change of zero energy DOS by a
magnetic field $H\propto q^2$; and (c) Low temperature superfluid
density.}
\end{figure}
\textit{Volovik effect and superfluid density} In this section, we
discuss the physical properties of various SC states, especially
the DOS, Volovik effect and superfluid density. We did the
calculations for all SC states listed in Tab.I. Since we found
that the qualitative behavior is the same for SC states belong to
the same category (full-gap, nodal-gap or gapless), we only
present the results for representative states, namely, the
full-gap SC state No.1, the nodal-gap SC state No.5 and gapless SC
state No.11. Fig.\ref{dos}(a) shows the DOS. The low energy
U-shaped (V-shaped) DOS in the full-gap (nodal-gap) SC states are
conventional. The gapless case is rather exotic. The low energy
DOS is not gapped but slightly piled up by SC order, due to the
presence of BdG FS. Fig.\ref{dos}(b) shows the effect of a
magnetic field on the DOS, which is calculated by a semi-classical
method as follows.\cite{Volovik} The effect of vortices is
simulated by averaging over the direction of the superflow
momentum $\mathbf{q}$ with $q\propto\sqrt{H}$. The change in zero
energy DOS $\Delta N(0,q)=\langle N(0,{\bf q})\rangle-N(0,0)$ can
be probed by the magnetic field dependent specific
heat.\cite{specific heat} It is seen in Fig.\ref{dos}(b) that the
Volovik effect is absent in the full-gap SC state, but is manifest
as a linear rise of $\Delta N$ with $q$ in the nodal-gap SC state.
Moreover, for the gapless state, the presence of magnetic field
actually reduces the DOS (as seen for moderate $q$).
Fig.\ref{dos}(c) shows the temperature dependence of superfluid
density. In the low temperature limit we neglect the temperature
dependence of the pairing amplitude. The full-gap and nodal-gap SC
states exhibit activated and linear drop, respectively.
Contrarily, the superfluid density is negative at all temperatures
in the gapless case. This arises from the fact that piling up of
low energy DOS leads to an overwhelming paramagnetic contribution
against the diamagnetic part. It indicates that the gapless state
is unstable upon phase twisting, implying a tendency toward
possible Fulde-Ferrell-Larkin-Ovchinnikov state or magnetic
ordering.

\textit{Discussion} Basing upon our analysis presented above and
recent experiments, we could discuss the possible pairing symmetry
of iron-based superconductors. Specific heat\cite{specific heat},
NMR\cite{NMR} penetrate length\cite{superfluid} and tunneling
\cite{Andreev} measurements are consistent with spin-singlet nodal
pairing with sign changes. On the other hand, on-site singlet
pairing is unfavorable to Hund's rule coupling, while bond-wise
triplet pairing is inconsistent with the antiferromagnetic
exchange.\cite{AF} We are then left with the bond-wise singlet
pairing cases No.3,4,6,7,and 9 in Tab.I. Out of these cases we
observe that only the $A_{2g}$ case No.4 and the $B_{2g}$ case
No.9 have nodal points in the $x$- and $y$-directions, which are
more relevant if the electron pockets are important. We propose
that Raman scattering and phase-sensitive probes could further
reduce the redundancy.

Some remarks are in order before closing. First, it should be
emphasized that the symmetry classification is robust, while the
assignment of a particular pairing symmetry has to be sharpened or
even altered according to further systematic and intrinsic
experimental results. Second, a general pairing matrix can be
decomposed into a linear combination of the bases. In principle,
bases belonging to different irreducible representations do not
mix, but mixing in other cases can not be ruled out on symmetry
ground alone. In particular, the two-dimensional representations,
like the case No.10 may be mixed into one of the
chiral-symmetry-breaking states shown in No.10'. Third, even if
the effect of As ions neglected so far are reconsidered, the point
group is still $D_{4h}$, so that the above symmetry classification
still holds. Finally, several authors propose models including
$d_{xy}$ or even all five d-orbitals.\cite{moreorbital} These may
be included in our analysis and result in more pairing bases. For
example $xy$ can pair with $yz$, which transform as $x$, forming
one component of the $E$ representations.

We became aware of a related work\cite{daisymm} after we finished
the present paper.

\begin{acknowledgments}
\textit{Acknowledgements} We thank Feng Zhou for technical help in
computation, and Zheng-Yu Weng for helpful discussions. QH thanks
The Center of Advanced Study of Tsinghua University where the
paper was finalized. The work was supported by NSFC 10325416, the
Ministry of Science and Technology of China (under the Grant No.
2006CB921802 and 2006CB601002) and the 111 Project (under the
Grant No. B07026).

\end{acknowledgments}

\end{document}